\documentclass[aps, prapplied, reprint, longbibliography]{revtex4-2}
\usepackage{xr-hyper, hyperref, graphicx, amsmath, amssymb, xcolor,lineno}
\graphicspath{{figures/}}

\begin{document}
\title{Perfect absorption by metal-contacted two-dimensional systems with ultra-proximate reflectors}

\author{Kirill Kapralov$^1$, Vladislav Atlasov$^1$, Alina Khisameeva$^2$, Viacheslav Muravev$^{3}$, Dmitry Svintsov$^1$}
\email[]{svintcov.da@mipt.ru}
\affiliation{$^1$Moscow Institute of Physics and Technology, Dolgoprudny 141700, Russia}
\affiliation{$^2$Institute of Solid State Physics, Vienna University of Technology, 1040 Vienna, Austria}
\affiliation{$^3$Institute of Solid State Physics RAS, Chernogolovka 142432, Russia}

\begin{abstract}
Electromagnetic absorbance by most two-dimensional electron systems is typically well below unity, which hinders both practical applications in photodetection and fundamental studies of their optical properties. Here, we show that a periodic structure comprised of narrow two-dimensional sections connected with wide perfectly conducting metal sections enables large absorbance. It reaches 50 \% provided the filling factor by the two-dimensional system $f$ equals its dimensionless conductivity $\eta=\sigma Z_0/2$, where $Z_0$ is the free-space impedance. The absorbance is further raised to 100 \% if the periodic structure is placed above a perfectly conducting electromagnetic reflector, and provided $f=2\eta$. Surprisingly, the optimal distance between two-dimensional system and reflector may fall well below the quarter of incident wavelength $\lambda_0/4$, which was assumed as conventional absorption enhancement condition in optics. For low filling factors $f\ll1$, large dielectric constants of the substrate, and grating periods comparable with $\lambda_0$, the optimal distance to reflector tends to zero. Above the critical values of the grating geometrical parameters, the absorbance maximum ceases to exist. The critical behavior manifests as a large-amplitude resonance in 'dirty' two-dimensional system with purely real conductivity, while enhancement of carrier momentum relaxation time lowers the resonant peak. Such resonance mimics the plasmonic one, but does not rely on high electron mobility.    
\end{abstract}
\maketitle

\section{Introduction} 

Absorbance of thin films and two-dimensional (2d) semiconductor structures is the central quantity for electromagnetic detectors and fundamental spectroscopic studies. The magnitude of absorption coefficient $\alpha$ allows multiplicative decomposition into 'material' and 'electromagnetic' factors:
\begin{equation}
\label{eq-absorbance-simple}
\alpha = \alpha_0 \times\frac{|{\bf E}|^2}{|{\bf E}_0|^2} ,  
\end{equation}
where ${\bf E}$ is the local electric field at the film surface, ${\bf E}_0$ is the field in the incident wave, and $\alpha_0={\rm Re}\sigma Z_0$ is the material-specific absorbance. The latter is comprised of the dynamic sheet conductivity of the film ${\rm Re}\sigma$ and the free-space impedance $Z_0$ equal to $4\pi/c$ in Gaussian units and 377 $\Omega$ in the SI units.

The 'material' factor of absorbance, $\alpha_0$, is well below unity for many 2d electron systems (2DES) and 2d materials, which represents a major obstacle to their optoelectronic applications and fundamental spectroscopic studies of their properties. As most prominent example, graphene displays $\alpha_0 = 2.3$ \% in the range of allowed interband transitions~\cite{Mak_graphene_conductivity}. Recent studies have revealed comparable values of absorption coefficient by other 2d materials~\cite{Absorbance_quantum_InAs,Stauber_absorption_2DMs}, while prior works have reported $\alpha_0$ order of several percent for mercury cadmium telluride quantum wells~\cite{Cesar_HgTe1,DePaula_HgTe2}. On the other side of electromagnetic spectrum, i.e. in the terahertz range, the absorbance $\alpha_0$ is governed by free-carriers~\cite{Zhukova_THzGraphene,Ikonnikov_JETP_THzHgTe}. Remarkably, it is again often well below unity. Indeed, the direct current resistivity of narrow-gap two-dimensional systems at charge neutrality (such as graphene and HgTe quantum wells) is order of $\rho_{\rm dc}\approx 10...50$ k$\Omega$~\cite{Kvon_2dSemimetal}. This translates into the low-frequency absorbance of $\alpha_0 \sim 1...4$ \%. An exclusion from the rule of small $\alpha_0$ is represented by quantum wells with large carrier density and long momentum relaxation time, where ${\rm Re}\sigma$ can be comparable with $Z_0^{-1}$~\cite{Muravev_relativistic_new,Muravev_relativistic_old}.

The presence of 'electromagnetic factor' $|{\bf E/\bf E}_0|^2$ can be both helpful or deteriorative for the total absorbance, depending on dielectric environment. For a thin film in uniform dielectric, $|{\bf E/\bf E}_0|^2 = |1 + \sigma Z_0/2|^{-2}$, which results in 'matched' absorbance of $\alpha = 50$ \% provided $\sigma Z_0/2 = 1$. This fact is widely used for optimization of thin metal film bolometers~\cite{Matched_IR_bolometer}. This kind of matching is not achievable for 2DES with $\alpha_0 \ll 1$. Further on, the absorbance by 2DES becomes even lower in electromagnetic experiments with controllable carrier density. Such control requires a highly conductive gate beneath the 2DES at distances $d$ order of hundreds of nanometers ~\cite{Yahniuk_THz_photocurrent,Olbrich_Thz_ratchet,Ma_photocurrent_probe,KrishnaKumar_photocurrent_probe}. Reflection of radiation from the gate is accompanied by the $\pi$ phase shift, which results in interference-type suppression of the electromagnetic factor $|{\bf E/\bf E}_0|^2 \ll 1$. Of course, in the visible and near-infrared ranges, the interference enhancement can appear instead for $d = \lambda_0/4n_{\rm sub}$, where $n_{\rm sub}$ is the refractive index of the substrate and $\lambda_0$ is the free-space wavelength~\cite{Cavity_enhanced_LWIR_PtSe2}. This principle of enhanced absorption dates back to Salisbury~\cite{salisbury_1,salisbury_2}, in fact, makes graphene monolayers visible~\cite{Making_graphene_visible}. However, such enhancement is hardly possible in the far-infrared and terahertz ranges, where $\lambda_0/4 \approx 1 ... 100$ $\mu$m exceeds the achievable thicknesses of gate dielectrics.

Strategies for electromagnetic enhancement of absorbance include vertical Fabry-Perot cavities~\cite{Cavity_enhanced_LWIR_PtSe2}, lateral plasmonic gratings~\cite{Perfect_absorption_grating}, and complex-shaped metasurfaces above (or directly atop) the 2DES~\cite{Thz_modulator_MS,MIR_Modulator,MIR_modulator_fast}. Most of these solutions allow one to raise the absorbance in a resonant fashion, i.e. in a narrow vicinity of the central frequency. There exists no general rule for design of light-concentrating metasurfaces, and their engineering represents a complex task involving numerical optimization algorithms~\cite{Lukianov_antennas}, often assisted by machine learning~\cite{Parmar_ML_absorbance}.

\begin{figure}[ht!]
    \includegraphics[width=1.0\linewidth]{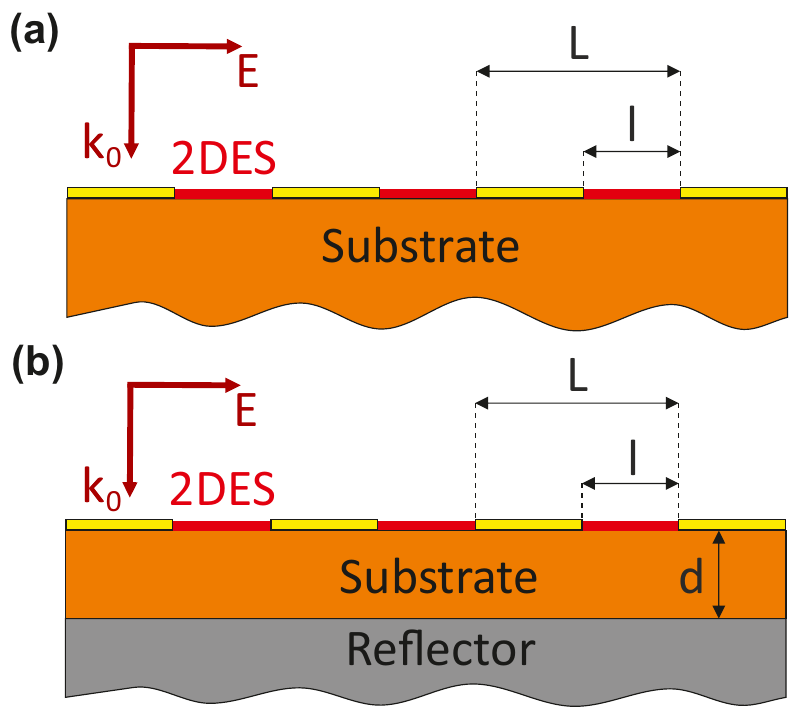}
    \caption{{\bf Structures of metal-2DES absorbers.}
One-dimensional periodic structure comprising 2DES of length $l$ and metal contacts of length $L$, illuminated by normally incident electromagnetic wave without (a) and with (b) a perfectly conducting reflector separated from the 2DES plane by a dielectric substrate of thickness $d$.
}
\label{fig_structure}
\end{figure}

Here, we demonstrate theoretically the possibility of large broadband absorption in 2DES with periodic perfectly conducting metal contacts. The structure is shown in Fig.~\ref{fig_structure}. We show that the incident field is expelled from metals and concentrated in weakly conducting 2DES. The absorption is maximized provided the filling factor $f$ of the periodic structure, i.e. the ratio of 2DES length $l$ to the period $L$, $f=l/L$, equals the real part of the dimensionless 2DES conductivity $\eta' = \sigma' Z_0/2$. For periodic structure in a uniform dielectric medium and purely real 2DES conductivity, the maximum absorbance reaches 50 \%. Further on, if the structure is placed at the distance $d$ above a perfectly conducting reflector, the maximum absorbance is raised to 100 \% and occurs provided $2\eta' = l/L$. Most surprisingly, the maximization can occur for ultra-small distances between 2DES and reflector, $d \ll \lambda_0/4n_{\rm sub}$, where $n_{\rm sub}$ is the refractive index of the substrate. Physical origin of the matching for ultra-small $d$ lies in the large phase shift acquired by the field transmitted through the grating with capacitive impedance $\Delta \Phi_c$, which adds up to the geometric phase shift of the wave acquired between 2DES and the reflector. Maximization of absorption with thin spacer layers can largely assist the terahertz optoelectronic experiments with gate-controlled 2DES. 

Several related proposals of absorption enhancement should be mentioned prior to our discussion. Jadidi et.al.~\cite{Jadidi_MGM_plasmons} have theoretically shown the possibility of 50 \% absorbance in metal-graphene lateral periodic structure, yet only in the plasmon-resonant regime. Wang and Tretyakov~\cite{Tretyakov_max_absorption} have proposed a two-dimensional metasurface with wide metal pads and narrow slits with graphene, derived the similar absorption maximization conditions, but did not reveal possible perfect absorption for deep-subwavelength distances to the reflector. Chen et.al.~\cite{Chen_experiment_MS} have experimentally realized such metasurface and demonstrated the perfect absorption experimentally using the reflector placed at a 'conventional' distance of $d = \lambda_0/4n_{\rm sub}$. Kuznetsov {\it et.al.} have implemented ultra-thin absorbers of similar type using absorbing dielectrics, not 2DES~\cite{Kuznetsov2012,Kuznetsov2016}. Studies of metal-based metasurfaces date back to Brown~\cite{Brown1,Brown2}, who has shown the possibility of achieving inductive- or capacitive-type impedance in a series of metal strips, depending on electric field orientation. Pendry {\it et.al.} re-addressed the inductive impedance of thin metal wires~\cite{Pendry1} and demonstrated a low-frequency analogue of plasmons in a wire medium~\cite{Pendry2}. The inductive character of the strip metasurface was used in~\cite{Muravev_resonator} for design of Fabry-Perot type cavity at sub-THz frequencies.  

\section{Theoretical model}

The structure under consideration is shown in Fig.~\ref{fig_structure}. It represents a one-dimensional periodic arrangement of 2DES sections of length $l$ and perfectly conducting metal contacts of length $L-l$, $L$ is the spatial period. The structure is placed above the substrate with permittivity $\varepsilon_{\rm sub}$, a perfectly conducting gate can be placed at distance $d$ beneath the grating. The electromagnetic wave is incident normally on the structure, its electric field is perpendicular to the metal contacts. Choosing the plane of 2d structure as $z=0$, directing the $x$ axis normally to the contacts, we write the incident field as ${\bf E}_{\rm inc} = {\bf e}_x E_0 e^{-i\omega t}$. 
 
The calculation of electromagnetic absorption is based on semi-analytical solution of the Maxwell's equations. Both 2DES, metal grid and reflector modify the incident field, such that the local field $E(x)$ in the 2DES plane ($z=0$) is largely different from $E_0$. To find $E(x)$, we first link its Fourier transform $E(q)$ to the Fourier transform of the surface current $J(q)$ in the 2DES:
\begin{equation}
  Z_0 J\left( q \right)=g(q)\left[ E\left( q \right)-{\tilde{E}_{0}}\left( q \right) \right],  
\end{equation}
where $\tilde E_0\left( q \right)$ is the Fourier transformed field in the top plane of the substrate in the absence of 2DES, and $g(q)$ is the fundamental solution of the two-dimensional wave equation in the Fourier representation. It is found by Fourier transform of the wave equation with respect to $x$-coordinate, and solution of the resulting second-order differential equation $E(q,z)$ with respect to the $z$-coordinate. The result for $g(q)$ in structure of Fig.~\ref{fig_structure} (b) reads as
\begin{equation}
\label{eq-fund-sol}
    g(q)=i{{k}_{0}}\left( \frac{1}{\kappa_1(q)}+\frac{\varepsilon}{\kappa_\varepsilon(q)}\frac{1+e^{-2\kappa_\varepsilon d}}{1-e^{-2\kappa_\varepsilon d}} \right),
\end{equation}
where $\kappa_\varepsilon \left( q \right)={{\left[ {{q}^{2}}-\varepsilon k_{0}^{2} \right]}^{1/2}}$ is the decay constant of the electromagnetic field outside the 2DES. The first term of (\ref{eq-fund-sol}) can be attributed to the field propagation in the medium above the 2DES, while the second -- to the field propagation into the substrate with dielectric constant $\varepsilon_{\rm sub}$. For 'radiative' wave vectors $q < \sqrt{\varepsilon_{\rm sub}} k_0$, the decay constant turns to the propagation constant with the sign convention $\kappa_\varepsilon \left( q \right) \rightarrow -i \sqrt{\varepsilon k_0^2-q^2}$.

As the link between Fourier components of current $J(q)$ and field $E(q)$ is established, there are two ways of passing to the real-space formulation. In one case~\cite{Fateev_transformation}, $E(x)$ can be presented as the convolution of currents $J(x')$ at all other spatial positions, the convolution kernel being the inverse Fourier transform of $g^{-1}(q)$. In the second case, the current $J(x)$ is written as a convolution of $E(x')$ at positions $x'$; the kernel of the convolution is the inverse Fourier of $g(q)$. The second method is preferable for structures including perfect conductors (PECs)~\cite{Tsymbalov_slot,Popov_EM_renormalization}. As the field inside PEC is zero, the convolution of electric field spans only over a small domain $x\in [-l/2; l/2]$. Physically, the local field $E(x)$ within 2DES uniquely determines the fields and currents in all surrounding space.  We further perform the inverse Fourier transform of (\ref{eq-fund-sol}) with the second method, and supplement the resulting integral equation with local Ohm's law $J(x) = \sigma E(x)$, thus arriving at
\begin{gather}
\label{eq-self-cons-coord}
Z_0 \sigma E\left( x \right)=\int\limits_{-L/2}^{L/2} dx \ g\left( x-{x}' \right)E\left( {{x}'} \right)+2{{E}_{0}},\\
g\left( x-{x}' \right)=\frac{1}{L} \sum\limits_{q_n} {g(q_n){{e}^{i{ q_n }\left( x-{x}' \right)}}},
\end{gather}
here $q_n=2\pi n/L$ is the set of reciprocal lattice vectors. In further analysis, we will consider two models of frequency dispersion: the Drude model relevant to low-frequencies (in the THz range and below) and the frequency-independent model relevant to graphene above the interband absorption edge. In principle, the solution of (\ref{eq-self-cons-coord}) is feasible for arbitrary model of conductivity dispersion $\sigma(\omega)$, i.e. at given radiation frequency $\sigma$ acts as an external numerical parameter.

We proceed to expand $E(x)$ in (\ref{eq-self-cons-coord}) into the orthogonal basis set of Legendre polynomials~\cite{Tsymbalov_slot}. Due to the inversion symmetry of the problem, only even polynomials contribute to the expansion:
\begin{equation}
\label{eq-Legendre}
E\left( x \right)={{E}_{0}}\sum\limits_{m=0}^{+\infty }{{{c}_{m}}{{P}_{2m}}\left( 2x/l \right)}.  
\end{equation}
Introducing the representation (\ref{eq-Legendre}) into (\ref{eq-self-cons-coord}) and evaluating the integrals over $x$ and $x'$ explicitly, we find the matrix representation of the scattering problem
\begin{equation}
 M_{km}c_m= \delta_{k0},
\end{equation}
\begin{multline}
  \label{eq-self-cons-Legendre}  
  {{M}_{km}}=\frac{{{\delta }_{mk}}}{4k+1}\eta +\frac{f}{2}\left[ i{{n}_{\rm sub}}\cot \left( {n}_{\rm sub}{{k}_{0}}d \right)+1 \right]{{\delta }_{k0}}{{\delta }_{m0}}-\\ - (-1)^{k+m} f \sum\limits_{n=1}^{+\infty }{g\left( {{q}_{n}} \right)  {j_{2k}}\left( \pi nf \right)j _{2m}\left( \pi nf \right) },
\end{multline}
where we have introduced the dimensionless 2DES conductivity $\eta = \sigma Z_0/2$, the refractive index of substrate $n_{\rm sub} = \varepsilon_{\rm sub}^{1/2}$, and the Kronecker delta symbol $\delta_{mk}$. The spherical Bessel functions of the order $k$, $j_k(\pi n f)$, appear in (\ref{eq-self-cons-Legendre}) upon transition from Fourier representation to the representation of Legendre polynomials $\int_{-1}^{1}{{{e}^{i y \xi }}{{P}_{k}}\left( \xi  \right)d\xi }=2{{i}^{k}}{{j}_{k}}\left( y \right)$.

Further solution should be performed numerically by truncating the matrix $M_{km}$ at some finite size $k_{\max}$. As the local field encoded by coefficients $c_m$ is found, $\alpha$ is readily computed with Joule's law for absorbed power and Pointing theorem for incident intensity:
\begin{equation}
\label{eq-absorbance}
    \alpha =2f{\eta }'\sum\limits_{n=0}^{\infty }{\frac{{{\left| {{c}_{n}} \right|}^{2}}}{2n+1}}.
\end{equation}
The solution of the linear system (\ref{eq-self-cons-Legendre}) converges rapidly even with a small size of the truncated matrix $k_{\rm max} = 1...10$. Instructively, approximation of the field by a single Legendre polynomial $P_0(2x/L) \equiv 1$, $E(x) = E_0 c_0$ is exact in the low-frequency limit. Indeed, the current cannot be accumulated or lost in the dc limit according to the continuity equation $\partial_t Q + \partial_x J = 0$; this enforces the local field to be simply a constant. The constant $c_0$, however, is not close to unity, i.e. the magnitude of the field is essentially modified by periodic structure. 

Retention of a single polynomial in the field expansion corresponds to the lumped circuit representation~\cite{Jadidi_MGM_plasmons,Tretyakov_max_absorption} of the scattering problem. To make this connection clearer, we multiply both sides of (\ref{eq-self-cons-Legendre}) by $2E_0/Z_0$ and retain the matrix element $M_{00}$ only. This results in
\begin{gather}
\label{eq-equiv-circuit}
    \frac{2E_0}{Z_0} = [\sigma + G_{\rm rad} - i \omega C_{\rm eff}] E,\\
    G_{\rm rad} = \frac{f}{Z_0}\left[ i{{n}_{\rm sub}}\cot \left( {n}_{\rm sub}{{k}_{0}}d \right)+1 \right],\\
    \label{eq-capacitance}
    C_{\rm eff} = 2f\varepsilon_0 \sum\limits_{n=1}^{+\infty }\left[\frac{1}{\kappa_1(q_n)}+\frac{\varepsilon}{\kappa_\varepsilon(q_n)}\frac{1+e^{-2\kappa_\varepsilon d}}{1-e^{-2\kappa_\varepsilon d}} \right]  {j_0^2}\left( \pi nf \right).
\end{gather}

An equivalent circuit representation immediately follows from (\ref{eq-equiv-circuit}). The source of electric current of magnitude $2E_0/Z_0$ feeds the 2DES with conductivity $\sigma$, the radiative conductance $G_{\rm rad}$, and simultaneously charges the capacitor $C_{\rm eff}$ formed between two metal pads. All conductance channels appear in parallel, forming an effective admittance $Y = \sigma + G_{\rm rad} - i \omega C_{\rm eff}$. The magnitude of drive current $2E_0/Z_0$ is exactly the sheet density of current in a perfectly conducting metal required to completely screen the incident field.

The equivalent circuit representation (\ref{eq-equiv-circuit}) is strictly derived from the spectrak approach to Maxwell's equations. Its advantage over equivalent circuit approach of~\cite{Tretyakov_max_absorption} lies in accurate definition of radiative conductance and effective capacitance $C_{\rm eff}$. We shall further see that $C_{\rm eff}$ is essentially modified by the reflector which, in turn, leads to electromagnetic matching with ultra-small distances $d$.

\begin{figure*}[ht!]
    \includegraphics[width=1.0\linewidth]{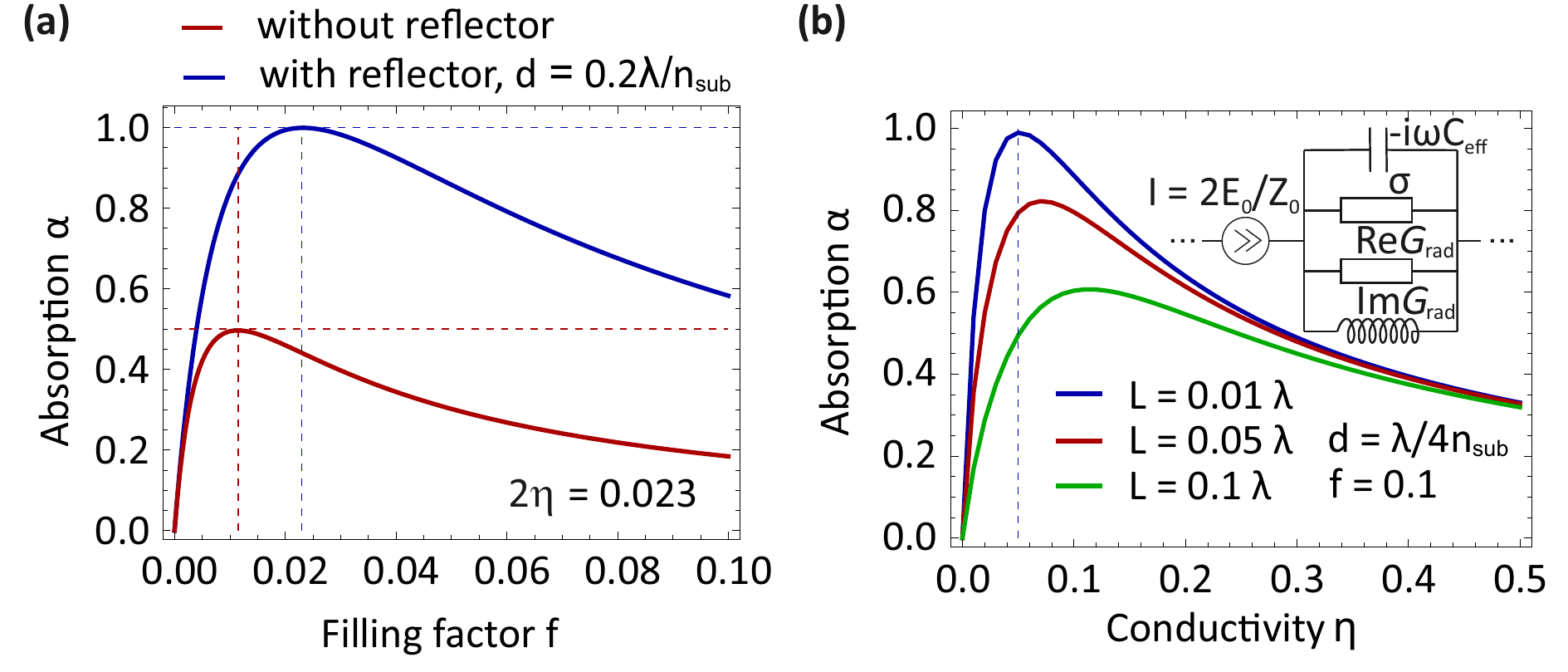}
    \caption{{\bf Perfect absorption in metal-contacted 2d electron systems.} 
(a) Absorption $\alpha$ as a function of the filling factor $f$ for the structure without a reflector with $n_{\mathrm{sub}}=1$ (red curve) and with a reflector with $n_{\mathrm{sub}}=2$ (blue curve). 
The horizontal dashed lines indicate the corresponding maximum absorption levels, while the vertical dashed lines mark the optimal filling factors: the red dashed line corresponds to $f=\eta$, and the blue dashed line corresponds to $f=2\eta$.
(b) Absorption $\alpha$ as a function of the conductivity $\eta$ for the structure with a reflector with $n_{\mathrm{sub}}=2$ for different grating periods. The vertical dashed line mark the optimal conductivity $\eta = f/2$. Inset: an equivalent circuit representation of the structure.
}
\label{fig_absorbance}
\end{figure*}

\section{Results: enhanced absorption of periodic metal-contacted 2DES}

\subsection{Broadband 50 \% absorption without reflectors}

The first remarkable property of one-dimensional metasurface including periodic metal and 2DES sections is its large absorbance in a broad range of frequencies without any plasmonic effects (i.e. for purely real 2DES conductivity $\eta = \eta'$) and back reflectors. This property appears already at the level of constant-field approximation and is further confirmed by the full numerical solution of the electromagnetic problem (\ref{eq-self-cons-Legendre}). To derive the absorption maximization conditions, we exclude the reflector from consideration by setting $d\rightarrow \infty$, $k_0 = k_0'+i\delta$ in (\ref{eq-equiv-circuit}), find the local field $E=c_0 E_0$, and introduce the result into the expression for absorbance (\ref{eq-absorbance}). This results in the following expression for absorbance
\begin{gather}
\label{eq-a-norefl}
\alpha \left( {{\eta }'} \right) \approx \frac{2f{\eta }'}{{{\left( {\eta }'+f \frac{n_{\rm sub}+1}{2} \right)}^{2}}+{{\left( 2f\frac{L}{{{\lambda }_{0}}}S \right)}^{2}}},\\
S =\sum\limits_{n=1}^{\infty }{\frac{{j^2_{0}}\left( \pi n f \right)}{\sqrt{{{n}^{2}}-{{\left( L/{{\lambda }_{0}} \right)}^{2}}}}},
\end{gather}
where we have introduced the dimensionless screening factor $S$ responsible for the inter-gap capacitance. {We observe that capacitive effects are negligible for absorption in subwavelength gratings with $L/\lambda_0 \ll 1$. Indeed, in that limit and for $f\ll 1$ we can estimate $S\approx3/2 - \ln (2\pi f)$. The large logarithmic factor of $-\ln (2\pi f)$ in capacitance cannot compensate the smallness of $L/\lambda_0$.}

The absorption maximization conditions, following from (\ref{eq-a-norefl}) for subwavelength gratings, become
\begin{equation}
\label{eq-a-max-norefl}
\eta'_{\rm opt} = f \frac{n_{\rm sub} + 1}{2},\qquad \alpha(\eta'_{\rm opt}) = \frac{1}{n_{\rm sub} + 1}.
\end{equation}
These conditions imply that 2DES with initially small material-limited absorbance $\alpha_0 \equiv 2\eta'\ll1$ can absorb up to 50 \% of the incident radiation if introduced into small gaps (with $f \sim \eta'$) between parallel metal contacts. We can speculate that electric potential drop across the unit cell $E_0 L$ is concentrated within a narrow 2DES-filled slit, such that local field $E\approx E_0 L/l$ is greatly enhanced. The enhancement is saturated at $f\sim\eta'$, which corresponds to the maximum of absorbance. In the transmission-line language, the effective load admittance seen by the incident wave is not $\sigma$, but rather $\sigma/f$ due to the field concentration by metal pads. This quantity readily reaches the free-space admittance $2/Z_0$, which ensures 50 \% absorbance. The dependence of absorbance on filling factor computed with (\ref{eq-a-norefl}) is shown in Fig.~\ref{fig_absorbance} (a) for $\varepsilon_{\rm sub}=1$; it indeed reaches $1/2$ for $f=\eta'$. 

The large absorbance (\ref{eq-a-max-norefl}), reaching 50 \% for equal refractive indices of the substrate and superstrate, is broadband in the following sense. It is maintained constant in the frequency range where the conductivity dispersion $\sigma(\omega)$ is negligible and the frequency is low, such that sub-wavelength condition $L/\lambda_0$ is maintained. Consider, for example, graphene at charge neutrality with $\rho_{dc} \approx 10$ k$\Omega$, the substrate being silicon with $n_{\rm sub}\approx 3.5$ in the terahertz range. The absorption maximization occurs at $f=8.5\times 10^{-3}$, $\alpha_{\rm max}$ reaches 22 \%. For grating period $L = 100$ $\mu$m ($l=850$ nm), large absorbance would persist up to wavelength $\lambda_0 \sim L = 100$ $\mu$m. In other words, large 22 \% absorbance would persist from nearly-zero frequencies up to $\omega/2\pi \sim 3$ THz.   

\subsection{Perfect absorption with deep-subwavelength distance to reflector} 

Though the metasurfaces of narrow 2DES-filled slits between metals enable the induction of large and broadband absorbance, they can still be improved in the following aspects. First, the maximum absorbance of $[n_{\rm sub}+1]^{-1}$ is still below unity, especially for optically dense substrates. Second, the requirement on slit width $f=\eta'$ can be very restrictive: already for the THz range and $\eta \sim 10^{-2}$, we face the need to fabricate slits with a width on the order of hundreds of nanometers.

The above problems are partially resolved by placement of a perfectly conducting screen at a distance $d$ below the 2DES-metal periodic structure. The screen need not to be metallic. In particular, highly doped silicon substrates used as bottom gates in experiments with 2d materials~\cite{Olbrich_Thz_ratchet,Karch_photon_drag_graphene} ensure nearly perfect reflection in the THz range. We proceed to study the absorbance by metal-2DES metasurfaces with back reflector, employing the uniform-field approximation (\ref{eq-equiv-circuit}) and supplementing our results with full numerical analysis.

Figure \ref{fig_absorbance} (b) shows the computed dependence of metasurface absorbance vs 2DES conductivity $\eta$ in the presence of a conventionally-placed reflector, $d = \lambda_0/4n_{\rm sub}$. The unit cell of the metasurface is subwavelength with $L/\lambda_0$ ranging from 0.01 to 0.1, the filling factor $f=0.1$, the refractive index of the substrate is $n_{\rm sub} = 2.0$. Three properties of this dependence are important. First, the absorbance maximum reaches 100 \% despite the substrate refractive index exceeding unity. Second, the maximum occurs provided $f = 2 \eta$, which is also independent of refractive index. Third, the optimal filling factor for mirror-backed metasurface is approximately twice that of a metasurface without mirror. It implies simpler fabrication tolerances for structures with reflecting gates. Increase in the grating period $L/\lambda_0$ reduces absorption. This trend is illustrated in Fig.~\ref{fig_absorbance} (b) with red and green lines.

\begin{figure*}[ht]
    \includegraphics[width=1\linewidth]{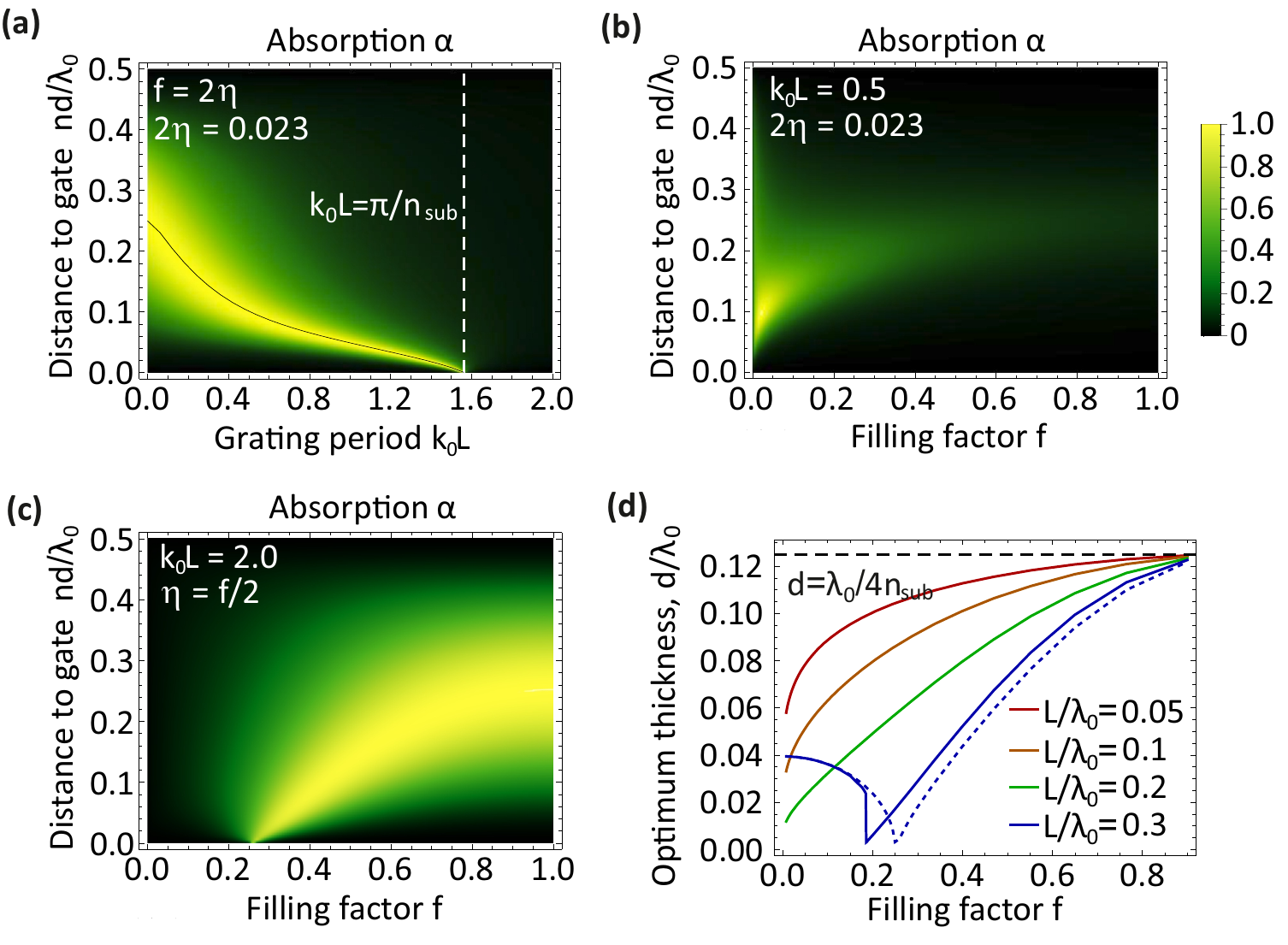}
    \caption{{\bf Perfect absorption with deep-subwavelength reflectors.} (a) Absorption $\alpha$ of the metal-2DES grating with a reflector vs the normalized grating period $k_0L$ and the normalized distance $n_{\mathrm{sub}}d/\lambda_0$ at the optimal filling factor $f=2\eta=0.023$. The white dashed line marks the limiting value $k_0L=\pi/n_{\mathrm{sub}}$, above which electromagnetic matching cannot be achieved. The black solid line shows the optimal $d$ yielding $100\%$ absorption (b) Absorption $\alpha$ vs the filling factor $f$ and the normalized distance $n_{\mathrm{sub}}d/\lambda_0$ computed for a fixed grating period $k_0L=0.5$ and conductivity $2\eta = 0.023$. (c) Absorption $\alpha$ vs the filling factor $f$ and the normalized distance $n_{\mathrm{sub}}d/\lambda_0$ for a large grating period $k_0L=2.0$ and optimal conductivity $\eta = f/2$. (d) Optimal distance $n_{\mathrm{sub}}d_{\rm opt}/\lambda_0$ providing maximum absorption vs the filling factor $f$ for different grating periods $k_0L$. At each point in (d), the conductivity $\eta$ is set to its optimum value. Solid lines are based on full numerical solution of (\ref{eq-self-cons-Legendre}), dashed line -- on the lumped approximation (\ref{eq-equiv-circuit})
}
\label{fig_cavity_absorbance}
\end{figure*}

All these observations are theoretically explained on the basis of lumped model (\ref{eq-equiv-circuit}) in the presence of back mirror. The amendment of the back mirror turns radiative conductivity into a complex quantity. For deep-subwavelength structures, the capacitive part of impedance is negligible. Absorption maximization thus requires:
\begin{equation}
{\rm Im} G_{\rm rad} = 0, \qquad \sigma_{\rm 2d} = {\rm Re} G_{\rm rad},
\end{equation}
which occurs precisely at
\begin{equation}
d = \frac{\lambda}{4n_{\rm sub}}, \qquad \eta' = f/2.
\end{equation}
Remarkably, for optimally placed reflector $d = \lambda/4n_{\rm sub}$, all information about substrate refractive index $n_{\rm sub}$ in the lumped model (\ref{eq-equiv-circuit}) is lost. This guarantees 100 \% absorbance independent of the substrate material.

The situation is more intriguing when capacitive contribution to the impedance of mirror-backed metasurface is non-negligible. The presence of finite $C_{\rm eff}$ yields negative contribution to the imaginary part of total admittance. The perfect absorption, in a general case, requires cancellation of the total reactance,
\(
\operatorname{Im}[G_{\rm rad}-i\omega C_{\rm eff}]=0,
\)
i.e. the compensation between the inductive response of the reflector-spacer region and the capacitive response of the metal–2DES grating. For conventionally-placed reflector at $d=\lambda_0/4n_{\rm sub}$, this simply reduces absorption, in agreement with trends in Fig.~\ref{fig_absorbance} (b). More generally, negative capacitive conductivity $- i \omega C_{\rm eff}$ competes with positive inductive conductivity ${\rm Im}G_{\rm rad}$ and shifts the optimum substrate thickness $d$ to lower values. Situations favoring the enhanced role of capacitive delay include large grating period in units of wavelength $L/\lambda_0$, low filling factor $f$ (in that case the capacitance diverges logarithmically, $C_{\rm eff} \sim - \ln (2 \pi f)$), and large permittivity $\varepsilon_{\rm sub}$ of the gate dielectric.

While the existence of perfect absorption in a reflector-backed conductive sheet is a known result~\cite{salisbury_1,salisbury_2}, the possibility of reducing $d$ below the quarter of radiation wavelength is non-trivial. The physical explanation of perfect absorption by capacitive metasurfaces at small $d$ lies in extra electromagnetic phase shift $\Delta\Phi_{c}$, acquired by electromagnetic radiation passing through the metal-2DES structure. This phase shift adds up to the geometric one $\Delta\Phi_g = 2\pi n_{\rm sub} d/\lambda_0$ and the one due to reflection from back mirror, $\Delta\Phi_{r} = \pi$. When $\Delta\Phi_{c}$ becomes comparable with $\Delta\Phi_{g}$, a pronounced reduction in optimal thickness $d$ is realized. Reduction in optimal $d$ can be also understood using a minimal model of uniform 2DES with artificially added imaginary part of conductivity ${\rm Im}\sigma<0$ (Supplementary Section I). In the same line, Ref.~\cite{Absorption-screened} has studied the absorption of screened 2DES described by Drude model and found, oppositely, $d > \lambda_0/4n_{\rm sub}$. The natural example of Drude model corresponds to the inductive conductivity, which increases $d$, while the artificial case of metal-contacted 2DES corresponds to the effective capacitive conductivity, which reduces $d$.

The perfect absorption at very low distances to the back reflector is illustrated in Fig.~\ref{fig_cavity_absorbance} (a,b), where we present the computed color maps of $\alpha$ at fixed 2d conductivity $\eta$, variable dielectric thickness $d$, and variable factors affecting the capacitance, $k_0L$ and $f$. Panel \ref{fig_cavity_absorbance}(b) shows the absorbance vs filling factor. At large values of $f$, the maximization of $\alpha$ occurs conventionally at $n_{\rm sub}d/\lambda_0=1/4$, while lowering of $f$ leads to a pronounced downward shift of the optimum thickness. The feasibility of perfect absorption at low optimum thickness is even more apparent from Fig.~\ref{fig_cavity_absorbance}(a), where the grating period $k_0L$ acts as a parameter. The optimum thickness decreases linearly with increasing $k_0L$, while after exceeding the critical value ($k_0L\approx 1.6$ in panel (a) of Fig.~\ref{fig_cavity_absorbance}) the absorption maximization is no more possible. {In terms of extra transmission phase $\Delta \Phi_c$, the critical point corresponds to $\Delta \Phi_c = \pi$. The geometric phase $\Delta \Phi_g$ required for matching, becomes formally negative in that case, which forbids the absorption maximization in all the range $d\in[0;\lambda_0/4n_{\rm sub}]$.} Black line in Fig.~\ref{fig_cavity_absorbance} (a) shows the optimum distance to the gate computed with lumped model (\ref{eq-equiv-circuit}). It agrees well with position of the absorption maximum from the full numerical solution of Eq.~(\ref{eq-self-cons-Legendre}). At large grating periods $k_0 L$, reduction in filling factor $f$ also results in a critical behavior. The fact is illustrated in Fig.~\ref{fig_cavity_absorbance} (c): absorption is maximized at $d\ll \lambda_0/n_{\rm sub}$ for $f\lesssim 0.4$, while for $f<0.25$ the absorption maximum disappears.

The trends toward small matching thickness are illustrated again in panel (d) of Fig.~\ref{fig_cavity_absorbance} in a compressed fashion. The panel shows the value of $d$ ensuring maximum absorbance vs filling factor at different $L/\lambda_0$ . The conductivity $\eta'$ at each point of these plots is set to the optimum value, ensuring maximum of $\alpha(\eta')$. The 'critical behavior' of $d$ at low filling factors and large grating periods is again clearly seen.  For relatively small grating periods [red to green lines in Fig.~\ref{fig_cavity_absorbance} (d)], the distance to reflector $d/\lambda_0$ ensuring 100 \% absorbance tends to zero as the filling factor $f$ tends to zero. Starting from relatively large grating periods [blue line in Fig.~\ref{fig_cavity_absorbance} (d), $L/\lambda_0 = 0.3$], the optimal distance $d$ shrinks to zero for some a finite value of $f$.


\begin{figure*}[ht!]
    \includegraphics[width=1\linewidth]{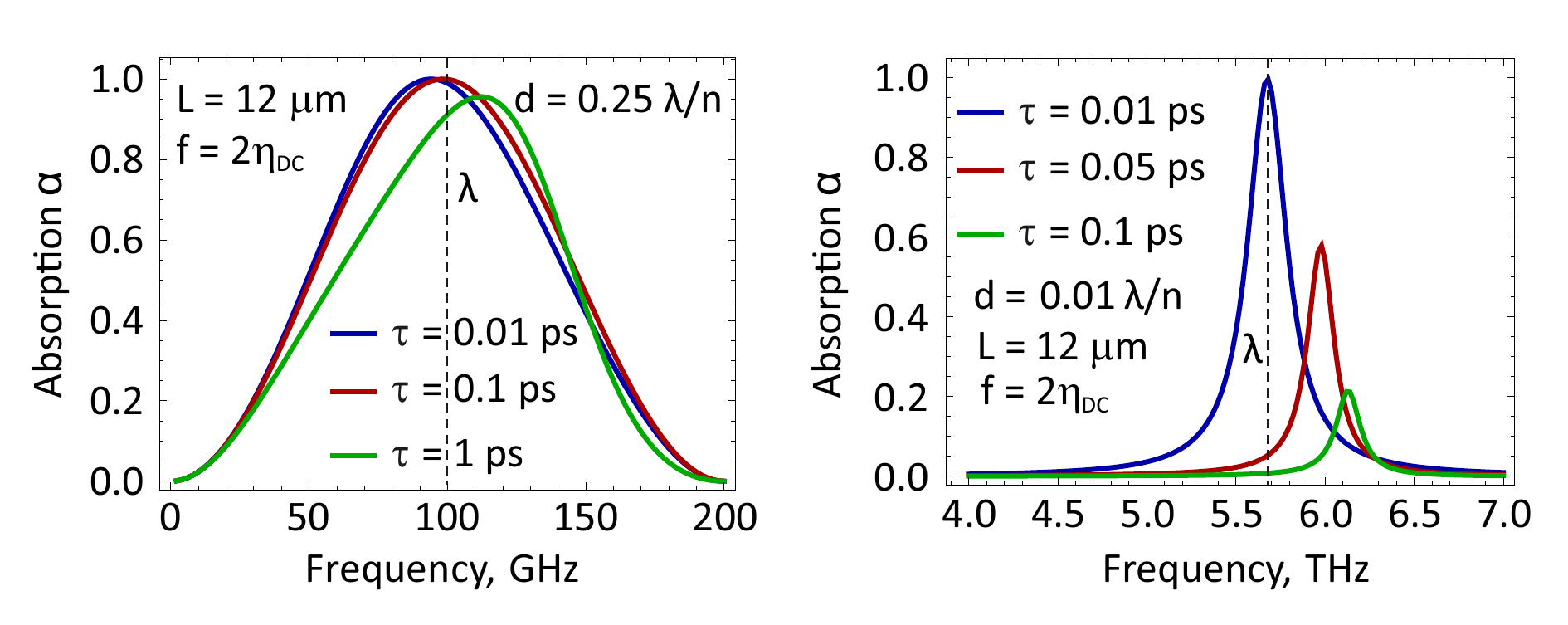}
    \caption{{\bf Resonant absorption in metal-2DES structures with a reflector.}  Absorption spectra calculated within the Drude model at carrier density $n_e=10^{12} \ \mathrm{cm}^{-2}$, effective mass $0.24 m_e$ and filling factor $f = 2 \eta(\omega=0)$ for different momentum relaxation times $\tau$ for a conventionally placed reflector at the distance $d=\lambda_{\text{res}}/(4n_{\mathrm{sub}})$ which equals $375~\mu \mathrm{m}$ (left) and for for an ultrathin substrate with $d=0.01\lambda_{\text{res}}/n_{\mathrm{sub}}$ which equals $265~ \mathrm{nm}$ (right). 
}
    \label{fig_freq_dependence}
\end{figure*}

These numerically observed criticalities can be predicted reliably with the lumped model (\ref{eq-equiv-circuit}). Indeed, the results obtained using large number of modes in Eq.~\ref{eq-self-cons-Legendre} agree well with constant-field approximation, as shown in Fig.~\ref{fig_cavity_absorbance} (d) with blue solid and dashed lines. As a result, the parameter combinations $\{d,f,L\}$ ensuring 100\% absorption can be estimated from the transcendental equation ${\rm Im}[G_{\rm rad} - i\omega C_{\rm eff}] = 0$. While its solution in a general case is not possible, we can find the critical value of $L$ ensuring the perfect absorption at $d\rightarrow 0$ and at small values of $f$ [the position of white dashed line in Fig.~\ref{fig_cavity_absorbance} (a)]. Performing the asymptotic estimate of the sum in Eq.~\ref{eq-capacitance} in the limit $d \rightarrow 0$, $f\rightarrow 0$, we find (see Supplementary Section II):
\begin{equation}
\label{eq-match-d}
    \frac{d_{\rm opt}}{\lambda_0/n_{\mathrm{sub}}}
    =
    \frac{
        \varepsilon \cot\left(\pi n_{\mathrm{sub}} L/\lambda_0\right)
    }{
        4(1+\varepsilon)
        \left[
            \frac{3}{2}
            -\ln\left(2\pi f \right)
        \right]}.
\end{equation}
As follows from Eq.~(\ref{eq-match-d}), the optimal distance to the reflector can be continuously reduced by increasing the grating period $k_0L$, and tends to zero as $k_0L$ approaches the limiting value $k_0L = \pi / n_{\mathrm{sub}}$. {The critical value of $f$ at fixed $L$ cannot be obtained with this simple method an requires a full solution of the transcendental matching equation.}

\subsection{Broadband and plasmon-mimicking absorption}

Our results for absorption have so far been presented as functions of dimensionless parameters: filling factor $f$, dimensionless period $L/\lambda_0$, and dimensionless conductivity $\eta$. From experimental viewpoint, it is more convenient to study the absorption spectra $\alpha(\omega)$. We proceed to show some unusual cases of spectral absorption for 2d metal-2DES gratings with back reflectors. The frequency dependence of conductivity is modeled with Drude formula $\sigma=n_e e^2\tau/[m(1-i\omega\tau)]$, where $n_e$ is the 2d electron density, $m$ is the effective mass, and $\tau$ is the momentum relaxation time. We intentionally set $\tau$ to be short, such that $\omega\tau\ll 1$ and plasmonic effects are excluded. All cases of resonant absorption presented below would result from competition of grating capacitance and phase shift acquired in back dielectric.

The spectrum of absorption for 'conventional' grating with not very small filling factor and period well below the wavelength is shown in Fig.~\ref{fig_freq_dependence} (a). The absorption maximum is broad and centered around $\omega = \pi c/(2n_{\rm sub} d)$. Increase in momentum relaxation time adds inductive contribution to the 2d conductivity and deteriorates the matching. 

A less conventional case is illustrated in Fig.~\ref{fig_freq_dependence} (b), where the distance to the reflector $d$ is intentionally taken as very small. The absorption $\alpha(\omega)$ develops a narrow peak resembling the plasmon resonance often observed in similar systems~\cite{Lee_Kang}. However, it has nothing to deal with plasmonics as $\omega\tau$ is below unity for $\tau =0.01$ ps (blue line), where the resonant absorption is the strongest. The origin of the presented resonance lies in passage through critical value of parameter $k_0L$ with increasing the frequency. A simple estimate shows that $\omega/2\pi\approx5.7$ THz corresponds to the nearly-zero optimum distance to the back reflector, while at larger frequencies the absorption maximization ceases to exist. Increase in momentum relaxation time, counter-intuitively, lowers the resonance, contrary to the case of plasmonic gratings in the non-retarded regime. The origin of lowering lies in inductive contribution to $\eta$, which competes with capacitance of the grating.

\section{Discussion and conclusions} 

The obtained results imply that the fundamental problem of low absorbance by 2d materials and numerous quantum well-based structures is resolved by amendment of periodic metal contacts with low filling factor. Another problem of interference-type suppression of absorption in 2d systems with highly conductive back gates is also resolved for such periodic metal-2DES arrangement. Precise selection of grating geometry enables interference type enhancement of absorption (instead of suppression) for arbitrarily small distances to the gate. The perfect absorption at small gate-2DES separation should appear in spectroscopic experiments as a narrow peak mimicking the plasma resonance, yet appearing for 'dirty' 2DES with purely real conductivity. Narrow absorption peaks in the dissipative regime were recently predicted in Ref.~\cite{Gorbenko_dissipative_PC} for yet another one-dimensional grating with gate-controlled alternating carrier density. Due to the approximate treatment of field modulation by the grating in \cite{Gorbenko_dissipative_PC}, it is not possible to reveal whether physical origins of these two effects are similar.

The considered 2DES-metal periodic structure with small filling factor not only enables perfect absorption, but is also beneficial for hot carrier effects~\cite{Hot_carriers}. The radiation-induced electron heating is proportional to the {\it absorbed} power density, which is a factor of $L/l$ larger than the {\it incident} power density. Increase in carrier temperature can be conveniently read out via bolometric effect. Strong zero-bias thermoelectric effect can appear in asymmetric gratings~\cite{Semkin2025}. Though their modeling goes far beyond the developed theory, we may suggest that perfect absorption at small filling factors and small distances to the gate would persist there as well.


\section*{Funding} 
The work was supported by the Ministry of Science and Higher Education of the Russian Federation, agreement \# 075-15-2025-608.

\section*{Conflict of interest} 
The authors of this work declare that they have no conflicts of interest

\bibliography{references}

\end{document}